\documentclass[runningheads]{llncs}
\usepackage[utf8]{inputenc} 
\usepackage[T1]{fontenc}
\usepackage{url}
\usepackage{ifthen}
\usepackage{cite}
\usepackage[cmex10]{amsmath} % Use the [cmex10] option to ensure complicance
\usepackage{amsmath}
\usepackage{multirow}
\usepackage{amsfonts}
\usepackage{graphicx}
\usepackage{algorithm}
\usepackage{graphicx}
\usepackage{latexsym}
\usepackage{amssymb}
\usepackage{amsfonts}
\usepackage{xcolor}
\usepackage{multirow}
\usepackage{algorithm}
\usepackage{enumerate}
\usepackage{caption}
\usepackage{mathtools}
\usepackage{enumitem}
\usepackage{subfig}
\usepackage{algpseudocode}% http://ctan.org/pkg/algorithmicx
\usepackage{etoolbox}
\usepackage{tikz}
\usetikzlibrary{tikzmark}
\usetikzlibrary{calc}
\usepackage{qtree}
\usepackage{tikz-qtree}
\usepackage{rotating}

\errorcontextlines\maxdimen

\newcommand{\ALGtikzmarkcolor}{black}% customise this, if you want
\newcommand{\ALGtikzmarkextraindent}{4pt}% customise this, if you want
\newcommand{\ALGtikzmarkverticaloffsetstart}{-.5ex}% customise this, if you want
\newcommand{\ALGtikzmarkverticaloffsetend}{-.5ex}% customise this, if you want
\makeatletter
\newcounter{ALG@tikzmark@tempcnta}

\newcommand\ALG@tikzmark@start{%
    \global\let\ALG@tikzmark@last\ALG@tikzmark@starttext%
    \expandafter\edef\csname ALG@tikzmark@\theALG@nested\endcsname{\theALG@tikzmark@tempcnta}%
    \tikzmark{ALG@tikzmark@start@\csname ALG@tikzmark@\theALG@nested\endcsname}%
    \addtocounter{ALG@tikzmark@tempcnta}{1}%
}

\def\ALG@tikzmark@starttext{start}
\newcommand\ALG@tikzmark@end{%
    \ifx\ALG@tikzmark@last\ALG@tikzmark@starttext
    \else
        \tikzmark{ALG@tikzmark@end@\csname ALG@tikzmark@\theALG@nested\endcsname}%
        \tikz[overlay,remember picture] \draw[\ALGtikzmarkcolor] let \p{S}=($(pic cs:ALG@tikzmark@start@\csname ALG@tikzmark@\theALG@nested\endcsname)+(\ALGtikzmarkextraindent,\ALGtikzmarkverticaloffsetstart)$), \p{E}=($(pic cs:ALG@tikzmark@end@\csname ALG@tikzmark@\theALG@nested\endcsname)+(\ALGtikzmarkextraindent,\ALGtikzmarkverticaloffsetend)$) in (\x{S},\y{S})--(\x{S},\y{E});%
    \fi
    \gdef\ALG@tikzmark@last{end}%
}

\apptocmd{\ALG@beginblock}{\ALG@tikzmark@start}{}{\errmessage{failed to patch}}
\pretocmd{\ALG@endblock}{\ALG@tikzmark@end}{}{\errmessage{failed to patch}}
\algblockdefx[Foreach]{Foreach}{EndForeach}[1]{\textbf{foreach} #1 \textbf{do}}{\textbf{end foreach}}

\DeclareGraphicsExtensions{.tif,.pdf,.jpeg,.png,.jpg,.bmp,.svg,.eps,.EMF}

\newcommand{\cM}{\mathcal{M}}
\newcommand{\cA}{\mathcal{A}}

\newcommand{\cT}{\mathcal{T}}

\newcommand{\cG}{\mathcal{G}}

\newcommand{\bX}{\mathbf{x}}
\newcommand{\bY}{\mathbf{y}}

\newcommand{\bV}{\mathbf{v}}

\newcommand{\bZ}{\mathbf{z}}

\newcommand{\poly}{\mathsf{poly}}
\newcommand{\decode}{\mathsf{decode}}
\newcommand{\diag}{\mathsf{diag}}

\newcommand{\supp}{\mathsf{supp}}
\newcommand{\test}{\mathsf{test}}

\newcommand{\plain}{\mathsf{plain}}
\newcommand{\Wx}{\mathsf{W}(x)}
\newcommand{\convert}{\mathsf{convert2NACGT}}

\newcommand{\sfA}{\mathsf{A}}
\newcommand{\sfB}{\mathsf{B}}
\newcommand{\sfW}{\mathsf{W}}

\begin{document}

\title{Efficient and error-tolerant schemes for non-adaptive complex group testing and \\its application in complex disease genetics}
%\vspace{-3mm}
% author names and affiliations
% use a multiple column layout for up to three different
% affiliations
%\begingroup
%\centering
%\endgroup
\author{Thach V. Bui\inst{1}, Minoru Kuribayashi\inst{3}, Mahdi Cheraghchi\inst{4} and Isao Echizen\inst{1,2}
%\vspace{-2mm}
\institute{SOKENDAI (The Graduate University for Advanced Studies), Kanagawa, Japan.
\and
National Institute of Informatics, Tokyo, Japan.\\
\email{\{bvthach,iechizen\}@nii.ac.jp}
\and
Okayama University, Okayama, Japan.\\
\email{ kminoru@okayama-u.ac.jp}
\and
Imperial College London, London, UK.\\
\email{m.cheraghchi@imperial.ac.uk}
}}

% make the title area
\maketitle

\begin{abstract}
The goal of combinatorial group testing is to efficiently identify up to $d$ defective items in a large population of $n$ items, where $d \ll n$. Defective items satisfy certain properties while the remaining items in the population do not. To efficiently identify defective items, a subset of items is pooled and then tested. In this work, we consider complex group testing (CmplxGT) in which a set of defective items consists of subsets of positive items (called \textit{positive complexes}). CmplxGT is classified into two categories: classical CmplxGT (CCmplxGT) and generalized CmplxGT (GCmplxGT). In CCmplxGT, the outcome of a test on a subset of items is positive if the subset contains at least one positive complex, and negative otherwise. In GCmplxGT, the outcome of a test on a subset of items is positive if the subset has a certain number of items of some positive complex, and negative otherwise.

For CCmplxGT, we present a scheme that efficiently identifies all positive complexes in time $t \times \poly(d, \ln{n})$ in the presence of erroneous outcomes, where $t$ is a predefined parameter. As $d \ll n$, this is significantly better than the currently best time of $\poly(t) \times O(n \ln{n})$. Moreover, in specific cases, the number of tests in our proposed scheme is smaller than previous work. For GCmplxGT, we present a scheme that efficiently identifies all positive complexes. These schemes are directly applicable in various areas such as complex disease genetics, molecular biology, and learning a hidden graph.% \textit{This paper is eligible for the best student paper award.}
\end{abstract}

\section{Introduction}
\label{sec:intro}

The task of combinatorial group testing is to efficiently identify up to $d$ defective items in a large population of $n$ items, where $d$ is usually much smaller than $n$. Defective items are the items that satisfy certain properties while the remaining items in the population, which are referred to as \textit{negative items}, do not. To efficiently identify the defective items, the items in a set are pooled into several (overlapping) subsets, and then the items in each subset are tested to determine whether they satisfy the properties. The outcome of a test on a subset is either positive, i.e., the subset satisfies the properties, or negative, i.e., the subset does not satisfy the properties. The procedure used to design the tests and obtain the outcomes is called ``encoding,'' and the procedure used to identify the defective items is called ``decoding.''

There are two main approaches to group testing: adaptive and non-adaptive. In adaptive testing, the design of a test depends on the designs of the previous tests. This approach is time-consuming for implementation. Non-adaptive group testing (NAGT) reduces the testing time because all tests are designed in advance and performed in parallel. The focus here is on NAGT.

In classical group testing (CGT)~\cite{dorfman1943detection}, the outcome of a test on a subset of items is positive if the subset has at least one defective item, and negative otherwise. The definition of a test outcome has been generalized to threshold group testing (TGT) with threshold $u$~\cite{damaschke2006threshold}, denoted as $u$-TGT, in which the outcome of a test on a subset is positive if the subset contains at least $u$ defective items, and negative otherwise.

The most sophisticated group testing is complex group testing (CmplxGT)~\cite{chang2010identification,chen2008upper,chin2013non}. CmplxGT originated in molecular biology~\cite{torney1999sets} and is also referred to as ``cover-free families''~\cite{stinson2004generalized} and ``learning hidden graphs''~\cite{angluin2006learning}. In classical CmplxGT (CCmplxGT)~\cite{torney1999sets}, an unknown set of subsets of items $D = \{D_1, \ldots, D_s \}$ is designated as a set of defective items in which each $D_a$ is called a positive complex, and the remaining items in the set $[n] \setminus \cup_{a = 1}^s D_a$ are designated as negative items, where $[n] = \{1, \ldots, n \}$. Suppose that $|D_a| \leq r$ and $\left| \cup_{a = 1}^s D_a \right| \leq d$. The outcome of a test on a subset of items is positive if the subset contains some positive complex $D_a$ and negative otherwise. Bui et al.~\cite{bui2018framework} recently introduced generalized CmplxGT (GCmplxGT) as follows. The outcome of a test on a subset of items is positive if the subset contains at least $u_a \leq u \leq r$ items from $D_a$ for some $a \in [s] = \{ 1, \ldots, s \}$ and negative otherwise.

Our objective is to minimize the number of tests and to efficiently identify defective items. For CGT, with non-adaptive design, the number of tests is $O(d^2 \ln{n})$~\cite{d1982bounds,du2000combinatorial,porat2008explicit,scarlett2016phase}. Several schemes~\cite{ngo2011efficiently,cheraghchi2013noise,cai2013grotesque,bui2018efficient} have been proposed for efficiently identifying defective items by using $O(d^{1 + o(1)} \ln{n})$ tests with decoding time $\poly(d, \ln{n})$. For TGT, most work has focused on the number of tests~\cite{damaschke2006threshold,cheraghchi2013improved,de2017subquadratic,chen2009nonadaptive} although work on the decoding procedure has recently proliferated~\cite{chen2009nonadaptive,chan2013stochastic,bui2018efficiently,reisizadeh2018sub}.

A matrix is an $(s, r; z]$-disjunct matrix if for any $s + r$ columns, there exist at least $z$ rows where each of the $r$ designated columns has 1s and each of the other $s$ columns has 0s. To tackle CCmplxGT, with non-adaptive design, we make use of the fact that the number of rows of an $(s, r; z]$-disjunct matrix is also the number of tests required in CCmplxGT. Chen et al.~\cite{chen2008upper} and Chin et al.~\cite{chin2013non} gave two upper bounds on the number of tests. Without considering erroneous outcomes, Abasi et al.~\cite{abasi2018non} reported the first algorithm requiring $O(t^{1 + o(1)} \ln{n})$ tests to identify positive complexes in time $\poly(t) \cdot O(n \ln{n})$, where $t$ is the number of rows in an $(s + r, r; 1]$-disjunct matrix. By considering erroneous outcomes, Abasi~\cite{abasi2018error} needed $t = \poly(s^r, \ln{n} )$ tests to identify all positive complexes in time $\poly(t) \cdot O(n \ln{n})$.

%D'yachkov et al.~\cite{d2016multistage} presented adaptive schemes that require at most $O(sr \ln{n})$ tests and stages. However, they are based on the assumption that the number of items goes to infinity, which is impractical.

%\vspace{-2mm}
\subsection{Contributions}
\label{sub:intro:contri}

To the best of our knowledge, this is the first work to focus on both the encoding and decoding procedures for GCmplxGT and to show the connection between CmplxGT and complex disease genetics (CDGs). The third contribution is the presentation of efficient encoding and decoding procedures for CCmplxGT with non-adaptive design in the presence of up to $\lfloor \frac{z-1}{2} \rfloor$ erroneous outcomes.

Let $h_0$ and $h_1$ be the numbers of rows in a $(d - r, r; z]$-disjunct matrix and a $(d - u, u; z]$-disjunct matrix. With CCmplxGT, all positive complexes can be identified using $h_0 \times O\left( \frac{d^2 \ln^4{n}}{\sfW^3(d\ln{n})} \right)$ tests in time $h_0 \times \poly(d, \ln{n})$, where $\Wx = \Theta \left( \ln{x} - \ln{\ln{x}} \right)$. With GCmplxGT, it takes $h_1 \times O\left( \frac{d^2 \ln^4{n}}{\sfW^3(d\ln{n})} \right)$ tests to identify all positive complexes in time $h_1 \times \poly(d, \ln{n}) + O(su^3dq^3)$, where $q = \sum_{a = 1}^s \binom{|D_a|}{u_a}$. Our results are directly applicable in various areas such as complex disease genetics, molecular biology, and learning a hidden graph.

%\vspace{-2mm}
\subsection{Comparison}
\label{sub:intro:cmp}

A detailed comparison with previous work is shown in Table~\ref{tbl:comparison} ($h_0$, $h_1$, and $q$ are defined in Section~\ref{sub:intro:contri}). Without considering the decoding procedure, Chen et al.~\cite{chen2008upper} showed that the number of non-adaptive tests is $O \left( z \left( \frac{p}{r} \right)^r \left( \frac{p}{s} \right)^s p \ln{\frac{n}{p}} \right)$, where $p = s + r$ and $\lfloor \frac{z-1}{2} \rfloor$ is the maximum number of erroneous outcomes. Abasi~\cite{abasi2018error} considered a fraction of the errors in test outcomes under the conditions $r < s$ and $r \leq O \left( \frac{\ln^2{p}}{\ln{\ln{p}}} \right)$. Abasi showed that all positive complexes can be identified in time $\poly(t) \times O(n \ln{n})$, where $t = O(r^{11} (4s)^{r + 7} \ln{n})$. Our proposed scheme has no constraints on either $r$ or $s$. We have shown that all positive complexes can be identified with $h_0 \times O\left( \frac{d^2 \ln^4{n}}{\sfW^3(d\ln{n})} \right)$ tests in time $h_0 \times \poly(d, \ln{n})$ with up to $\lfloor (z-1)/2 \rfloor$ erroneous outcomes. Our decoding time is thus better than Abasi's work once $n$ is large enough. Moreover, when $s > d$, the number of tests in our proposed scheme is also smaller than the one proposed by Abasi.

With GCmplxGT, all positive complexes can be identified in time $h_1 \cdot \poly(d, \ln{n}) + O(su^3dq^3)$ by using $h_1 \cdot O\left( \frac{d^2 \ln^4{n}}{\sfW^3(d\ln{n})} \right)$ tests with up to $\lfloor \frac{z-1}{2} \rfloor$ erroneous outcomes.

\begin{table*}[t]
\caption{Comparison with existing work. To simplify notation, we set $p = s + r; \ \sfA_0 = O \left( \frac{d^{3.57} \ln^{6.26}{n}}{\sfW^{6.26}(d \ln{n})} + \frac{d^6 \ln^4{n}}{\sfW^4(d \ln{n})} + \frac{d^2 \ln^3{n}}{\sfW^3(d \ln{n})} \right)$, and $\alpha = \Omega \left( \frac{1}{(r + 1) r^r \binom{r + s - 1}{r}} \right)$.}% Let $h_0$ and $h_1$ be the numbers of rows in a $(d - r, r; z]$-disjunct matrix and a $(d - u, u; z]$-disjunct matrix.} $q = \sum_{a = 1}^s \binom{|D_a|}{u_a}$,

\begin{center}
\scalebox{.93}{
\begin{tabular}{|c|c|c|c|c|c|}
\hline
& Scheme & \begin{tabular}{@{}c@{}} Conditions placed \\on $r$ and $s$ \end{tabular} & \begin{tabular}{@{}c@{}} Error \\tolerance \end{tabular} & \begin{tabular}{@{}c@{}} Number of tests \\ $t$ \end{tabular} & \begin{tabular}{@{}c@{}} Decoding \\complexity \end{tabular} \\
\hline
\multirow{4}{*}{\begin{sideways} \begin{tabular}{@{}c@{}} CCmplxGT \end{tabular} \end{sideways}} & \begin{tabular}{@{}c@{}} Abasi~\cite{abasi2018error} \end{tabular} & \begin{tabular}{@{}c@{}} $r < s$ \\ $r \leq O \left( \frac{\ln^2(r + s)}{\ln{\ln(r + s)}} \right)$ \end{tabular} & $\leq \alpha t$ & $O(r^{11} (4s)^{r + 7} \ln{n})$ & $\poly(t) \cdot O(n \ln{n})$ \\ \cline{2-6}
& \begin{tabular}{@{}c@{}} Chen et al.~\cite{chen2008upper} \end{tabular} & None & $\lfloor \frac{z - 1}{2} \rfloor$ & $O \left( z \left( \frac{p}{r} \right)^r \left( \frac{p}{s} \right)^s p \ln{\frac{n}{p}} \right)$ & Not available \\ \cline{2-6}
& \begin{tabular}{@{}c@{}} \textbf{Corollary~\ref{cor:NACCmplxGT:1}} \end{tabular} & None & $\lfloor \frac{z - 1}{2} \rfloor$ & $O \left( h_0 \times \frac{d^2\ln^2{n}}{\sfW^2(d\ln{n})} \right)$ & $h_0 \times \sfA_0$ \\ \cline{2-6}
& \begin{tabular}{@{}c@{}} \textbf{Corollary~\ref{cor:NACCmplxGT:2}} \end{tabular} & None & $\lfloor \frac{z - 1}{2} \rfloor$ & $O\left( h_0 \times \frac{d^2\ln^3{n}}{\sfW^2(d\ln{n})} \right)$ & $h_0 \times O\left( \frac{d^2 \ln^4{n}}{\sfW^3(d\ln{n})} \right)$ \\ \cline{2-6}
\hline
\multirow{2}{*}{\begin{sideways} \begin{tabular}{@{}c@{}} GCmplxGT \end{tabular} \end{sideways}} & \begin{tabular}{@{}c@{}} \\ \textbf{Corollary~\ref{cor:NAGCmplxGT:1}} \\ $\ $ \end{tabular} & None & $\lfloor \frac{z - 1}{2} \rfloor$ & $O \left( h_1 \times \frac{d^2\ln^2{n}}{\sfW^2(d\ln{n})} \right)$ & \begin{tabular}{@{}c@{}} $h_1 \times \sfA_0$ \\ $ + O(su^3dq^3)$ \end{tabular}  \\ \cline{2-6}
& \begin{tabular}{@{}c@{}} \\ \textbf{Corollary~\ref{cor:NAGCmplxGT:2}} \\ $\ $ \end{tabular} & None & $\lfloor \frac{z - 1}{2} \rfloor$ & $O\left( h_1 \times \frac{d^2\ln^3{n}}{\sfW^2(d\ln{n})} \right)$ & \begin{tabular}{@{}c@{}} $h_1 \times O\left( \frac{d^2 \ln^4{n}}{\sfW^3(d\ln{n})} \right)$ \\ $ + O(su^3dq^3)$ \end{tabular} \\ \cline{2-6}
\hline
\end{tabular}}
\end{center}

\label{tbl:comparison}
\end{table*}

%\vspace{-2mm}
\subsection{Applications}
\label{sub:apps}
%In this section, we outline two scenarios in which our contributions could play an important role and describe how we improve on earlier work.

\subsubsection{Complex disease genetics}

Complex diseases, e.g., Alzheimer's disease and Parkinson's disease, are caused by a combination of genetics and other factors, most of which have not been identified~\cite{craig2008complex}. This work addresses only CDGs~\cite{schork1997genetics} that are caused by a combination of many genes. Our goal is to efficiently identify those genes via biological data such as protein-protein interaction (PPI).

Let us denote a gene contributing to a disease as a defective gene. There are several genes contributing to the disease, though the number of non-defective genes outnumbers of the number of defective genes. Let us call a set of these genes a positive complex and a set of genes a complex. A complex may contain a positive complex and there might be more than one positive complex. The outcome of a PPI test on a complex is positive, i.e., the disease occurs, if there are a certain number of defective genes in a positive complex jointly appearing in the complex, and negative otherwise. The problem turns into CmplxGT and is thus resolvable with our proposed scheme.

\vspace{-3mm}
\subsubsection{Molecular biology}

Torney~\cite{torney1999sets} introduced a problem in molecular biology as follows. Consider a set $N$ of $n$ molecules. Let $D = \{D_1, \ldots, D_s \}$ be an unknown set of subsets of molecules to be identified. The molecules in each subset $D_a$ cause a certain biological phenomenon. A subset of $N$ is called \textit{a complex}, and each complex $D_a$ is called \textit{a positive complex}. An experiment conducted for any subset of $N$ has two possible outcomes: ``positive'' if the subset contains at least one $D_a$, and ``negative'' otherwise. Our goal is to identify each positive complex $D_a$ such that the number of experiments is as small as possible and the processing time is as short as possible.

The definition of Torney's problem is identical to the CCmplxGT problem and can be generalized as follows. A certain biological phenomenon occurs if a certain number of molecules in some $D_a$ jointly appear. This generalization can be viewed as GCmplxGT and is thus resolvable with our proposed scheme.

\vspace{-3mm}
\subsubsection{Learning a hidden hypergraph}

Angluin and Chen~\cite{angluin2006learning} described the problem of learning a hidden hypergraph as follows. Consider a set of $n$ items. The objective is to identify an unknown family $D = \{ D_1, \ldots, D_s \}$ from the given family $C$ of subsets of $[n]$. The family $C$ is viewed as a hypergraph with the vertex set containing $n$ items. Every $D_a$ is considered to be an edge of the hypergraph, and $D$ is a hidden graph. The only operation to be carried out is to test whether a subset of $n$ vertices contains an edge of $D$. Precisely, the outcome of a test on a subset of items is positive if the subset contains all members of some $D_a$, and negative otherwise. The goal is to identify the hidden subgraph $D$ in the given hypergraph $C$ with the minimum number of tests. This problem turns into a CCmplxGT problem and is thus resolvable with our proposed scheme.

\vspace{-2mm}
\section{Preliminaries}
\label{sec:pre}
\vspace{-2mm}
\subsection{Notation}
\label{sub:pre:notations}

A multiset, denoted with a capital letter with an ``$\star$'', is a set that allows multiple instances of its elements. A plain set, denoted with a capital letter with a ``$\prime$'', is a set containing indivisible elements only. Function $\plain(\cdot)$ creates a plain set by taking all indivisible elements in the input set. For example, $A^\star = \{1, 2, 2 \}$ is a multiset, $D = \{ D_1, D_2 \} = \{ \{1, 4 \}, \{ 4, 8\} \}$ is a set, and $D^\prime = \plain(D) = D_1 \cup D_2 = \{ 1, 4, 8 \}$ is a plain set.

For consistency, we use capital calligraphic letters for matrices, non-capital letters for scalars, and bold letters for vectors. All matrices here are binary. The intersection of $l$ columns of a $t \times n$ matrix $\cT$ is defined as $\bigwedge_{i=1}^l \cT_{j_i} = \left( \bigwedge_{i = 1}^l m_{1 j_i}, \ldots, \bigwedge_{i=1}^l m_{t j_i} \right)^T$. Notation $[m]$ represents set $\{1, 2, \ldots, m \}$. 

Function $\diag(\cdot)$ is used to create a diagonal matrix constructed from the input vector. The support set for $\bV = (v_1, \ldots, v_w)$ is $\supp(\bV) = \{j \mid v_j \neq 0 \}$.

Let $D$ and $D^\prime = \plain(D)$ be the defective set consisting of positive complexes and the plain set of $D$. Parameters $n, d$, and $\bX = (x_1, \ldots, x_n)^T$ represent the number of items, the maximum number of defective items, and the representation vector of $n$ items. Finally, let $\cT_{i, *}, \cG_{i, *}, \cM_{i,*}$, and $\cM_j$ be row $i$ of matrix $\cT$, row $i$ of matrix $\cG$, row $i$ of matrix $\cM$, and column $j$ of matrix $\cM$, respectively.

%\vspace{-2mm}
\subsection{Measurement matrix}
\label{sub:pre:measure}
%\vspace{-1mm}
For vector $\bX = (x_1, \ldots, x_n)^T$, $x_j = 0$ means that item $j$ is negative, and $x_j \neq 0$ means that item $x_j$ is defective. Note that $D^\prime = \supp(\bX)$. For a $t \times n$ binary measurement matrix $\cT = (t_{ij})$, item $j$ is represented by column $\cT_j$ and test $i$ is represented by row $\cT_{i, *}$; $t_{ij} = 1$ if item $j$ belongs to test $i$, and $t_{ij} = 0$ otherwise.

Let $\test(S)$ be the test on subset $S \subseteq [n]$. The outcome of the test is either positive (1) or negative (0) and depends on the definition of $D$ and $S$. The non-adaptive tests on $n$ items using $\cT$ are defined as
\begin{align}
\mathbf{y} = \cT \bullet \mathbf{x} &= \begin{bmatrix}
\test \left( \supp(\cT_{1, *}) \cap \supp(\bX) \right) & \ldots & \test \left( \supp(\cT_{t, *}) \cap \supp(\bX) \right)
\end{bmatrix}^T \nonumber \\
&= \begin{bmatrix} y_1 & \ldots & y_t \end{bmatrix}^T,
\label{eqn:encMeasurement}
\end{align}
where $y_i = \test \left( \supp(\cT_{i, *}) \cap \supp(\bX) \right)$ is the outcome of test $i$ corresponding to row $\cT_{i, *}$, and $\bullet$ is the test operator. The procedure to obtain $\bY$ is called \textit{encoding}. The procedure to recover $\bX$ from $\bY$ and $\cT$ is called \textit{decoding}.

For CGT and TGT, notation $\bullet$ can be explicitly defined and vector $\bX$ is viewed as a binary vector in which $x_j = 1$ (resp., $x_j = 0$) means item $j$ is defective (resp., negative). With CGT, to avoid ambiguity, we change notation $\bullet$ to $\odot$ and use a $k \times n$ measurement matrix $\cM$ instead of the $t \times n$ matrix $\cT$. The outcome vector $\bY$ in~\eqref{eqn:encMeasurement} is equal to
\begin{equation}
\label{eqn:disjunct}
\bY = \cM \odot \bX = \begin{bmatrix}
\cM_{1, *} \odot \bX \\
\vdots \\
\cM_{k, *} \odot \bX
\end{bmatrix}
= \begin{bmatrix}
\bigvee_{j=1}^{n} x_j \wedge m_{1j} \\
\vdots \\
\bigvee_{j=1}^{n} x_j \wedge m_{kj}
\end{bmatrix} = \bigvee_{\substack{j=1, x_j = 1}}^{n} \cM_j = \begin{bmatrix}
y_1 \\
\vdots \\
y_k
\end{bmatrix},
\end{equation}
where $y_i = \cM_{i, *} \odot \bX = \bigvee_{j=1}^{n} x_j \wedge m_{ij} = \bigvee_{j=1, x_j = 1}^{n} m_{ij}$ for $i = 1, \ldots, k$.

With $u$-TGT, to avoid ambiguity, we change notation $\bullet$ to $\otimes_u$. Outcome vector $\bY$ in~\eqref{eqn:encMeasurement} is equal to $\bY = \cT \otimes_u \bX = [\cT_{1, *} \otimes_u \bX \ \ldots \ \cT_{t, *} \otimes_u \bX]^T = [y_1 \ \ldots \ y_t]^T$, where $y_i = \cT_{i, *} \otimes_u \bX = 1$ if $\sum_{j=1}^n x_j t_{ij} \geq u$, and $y_i = 0$ otherwise for $i \in [t]$. %= \cT_{i, *} \otimes_u \bX

\subsection{Disjunct matrices}
\label{sub:pre:disjunct}

Disjunct matrices were first introduced by Kautz and Singleton~\cite{kautz1964nonrandom} as \textit{superimposed codes} and then generalized by Stinson and Wei~\cite{stinson2004generalized} and D'yachkov et al.~\cite{d2002families}. The formal definition of a disjunct matrix is as follows.

\begin{definition}
An $m \times n$ binary matrix $\cT$ is called a $(d, r; z]$-disjunct matrix if, for any two disjoint subsets $S_1, S_2 \subset [n]$ such that $|S_1| = d$ and $|S_2| = r$, there exists at least $z$ rows in which there are all 1's among the columns in $S_2$ while all the columns in $S_1$ have 0's, i.e., $\left\vert \bigcap_{j \in S_2} \supp \left( \cT_j \right) \big\backslash \bigcup_{j \in S_1} \supp \left( \cT_j \right) \right\vert \geq z$.% Parameter $\lfloor (z-1)/2 \rfloor$ is usually referred to as the \textit{error tolerance}.
\label{def:threshDisjunct}
\end{definition}

Chen et al.~\cite{chen2008upper} gave an upper bound on the number of rows for $(d, u; z]$-disjunct matrices as follows.

\begin{theorem}~\cite[Theorem 3.2]{chen2008upper}
\label{thr:ChenUpper}
For any positive integers $d, u, z$, and $n$ with $p = d + u \leq n$, there exists a $t \times n$ $(d, u; z]$-disjunct matrix with $t = O \left( z \left( \frac{p}{u} \right)^u \left( \frac{p}{d} \right)^d p \ln{\frac{n}{p}} \right)$.
\end{theorem}

When $r = z = 1$, a $(d, r; z]$-disjunct matrix becomes a $d$-disjunct matrix. If $\cM$ is $d$-disjunct, vector $\bX$ can be recovered from $\bY = \cM \odot \bX$. Bui et al.~\cite{bui2018efficient} numerically showed that the number of tests in nonrandom construction (each column can be generated without using probability) is optimal for practice (albeit it is not good in term of complexity). Therefore, we use that result here. %We consider only $d$-disjunct matrices in which the columns can be computed in time $\poly(k)$.

\begin{theorem}~\cite[Theorem 8]{bui2018efficient}
\label{thr:mainNonrandom}
Let $1 \leq d \leq n$ be integers. Then there exists a nonrandom $d$-disjunct matrix $\cM$ with $k = O \left(\frac{d^2 \ln^2{n}}{\mathsf{W}^2(d\ln{n})} \right).$ Each column in $\cM$ can be computed in time $O(k^{1.5}/d^2)$, so matrix $\cM$ can be used to identify up to $d^\prime$ defective items, where $d^\prime \geq \left\lfloor \frac{d}{2} \right\rfloor + 1$, in time $O \left( \frac{d^{3.57} \ln^{6.26}{n}}{\mathsf{W}^{6.26}(d \ln{n})} \right) + O \left( \frac{d^6 \ln^4{n}}{\mathsf{W}^4(d \ln{n})} \right).$
\end{theorem}

The decoding complexity can be reduced by increasing the number of tests:

\begin{theorem}~\cite[Corollary 3]{bui2018efficient}
\label{thr:mainNonrandom2}
Let $1 \leq d \leq n$ be integers. There exists a nonrandom $k \times n$ measurement matrix $\cT$ with $k = O \left( \frac{d^2 \ln^3{n}}{\mathsf{W}^2(d \ln{n})} \right)$, which is used to identify at most $d$ defective items in time $O(k)$. Moreover, each column in $\cT$ can be computed in time $O \left(\frac{d \ln^4{n}}{\mathsf{W}^3(d\ln{n})} \right)$.
\end{theorem}

The procedure for obtaining $\bX$ from $\bY$ is denoted as $\supp(\bX) = \decode(\cM, \bY)$.

\section{Problem definitions}
\label{sec:def}

We classify CmplxGT into two categories: CCmplxGT and GCmplxGT. Our goal is to identify positive complexes as quickly as possible with a small number of tests. The formal definitions of CCmplxGT and GCmplxGT are given below.

\begin{definition}[Classical complex group testing]
\label{def:CCmplxGT}
Let $1 \leq r,s \leq d < n$ be integers. Consider a set $N$ of $n$ items. Suppose that $D = \{ D_1, \ldots, D_s \}$ is an unknown set of positive complexes of $N$, where $|D^\prime| = \left| \cup_{a = 1}^s D_a \right| \leq d$, $|D_a| \leq r$, and $D_a \not\subseteq D_b$ for $a \neq b \in [s]$. With CCmplxGT, the outcome of a test on a subset of $N$ is positive if the subset contains some $D_a$, and negative otherwise.
\end{definition}

We define generalized complex group testing as follows. %Each positive complex $D_a$ in CCmplxGT can be considered as a defective set in $|D_a|$-TGT.

\begin{definition}[Generalized complex group testing]
\label{def:GCmplxGT}
Let $1 \leq r,s \leq d < n$ be integers. Consider a set $N$ of $n$ items. Suppose that $D = \{ D_1, \ldots, D_s \}$ is an unknown set of positive complexes of $N$, where $|D^\prime| = \left| \cup_{a = 1}^s D_a \right| \leq d$, $|D_a| \leq r$, and $D_a \not\subseteq D_b$ for $a \neq b \in [s]$. Any $D_a$ is a defective set in $u_a$-TGT, where $0 < u_a \leq u \leq d$ for $a = 1, \ldots, s$. With GCmplxGT, the outcome of a test on a subset of $N$ is positive if the subset contains at least $u_a$ items in $D_a$ for some $a \in [s]$, and negative otherwise.
\end{definition}

It is obvious that when $u_a = |D_a|$ for every $a \in [s]$, Definition~\ref{def:GCmplxGT} reduces to Definition~\ref{def:CCmplxGT}. We then define a (incomplete) positive sub-complex as follows.

\begin{definition}
\label{def:subcomplex}
Consider generalized complex group testing as defined in Definition~\ref{def:GCmplxGT}. A set $I$ is called \textit{a positive sub-complex} if $|I \cap D_a| \geq u_a$ for some $a \in [s]$. Otherwise, $I$ is called \textit{an incomplete positive sub-complex}.
\end{definition}

For example, let $r = 4, s = 3, d = 7, n = 10$ and $N = \{1, 2, \ldots, 10 \}$. Suppose that $D = \{ D_1 = \{1, 2 \}, D_2 = \{2, 3, 4 \}, D_3 = \{1, 3, 7, 8 \} \}$ and $u_1 = 2, u_2 = 2$, and $u_3 = 3$. Then $I_1 = \{1, 2 \}$, $I_2 = \{2, 4 \}$, and $I_3 = \{3, 7, 8 \}$ are positive sub-complexes. A few incomplete positive sub-complexes are $I_4 = \{1, 5 \}, I_5 = \{3, 6, 7, 8 \}$ and $I_6 = \{ 1, 7, 8, 9, 10 \}$.

Because an incomplete positive sub-complex does not affect the outcome of a test on a set containing it, all items in it are considered to be \textit{negative} for that test. Note that some items in an incomplete defective sub-complex for a test could be defective items in other tests. For example, consider a subset $I_7 = \{1, 2, 3, 9, 10 \}$, and with the other settings as the same as in the preceding paragraph. The outcome of a test on $I_7$ is positive: items 1 and 2 are identified as defectives, whereas items 3, 9, and 10 are identified as negatives.

\section{Review of Bui et al.'s scheme}
\label{sec:review}

A part of the scheme proposed by Bui et al.~\cite{bui2018efficiently} is reviewed here. For $u$-TGT, Bui et al. considered a special case in which the number of defective items equals the threshold; i.e., $|\supp(\bX)| = u$. Let $\cM = (m_{ij})$ be a $k \times n$ $d$-disjunct matrix as described in Section~\ref{sub:pre:disjunct}. Then a measurement matrix is created as
\begin{equation}
\label{eqn:elementaryMatrix}
\cA = \begin{bmatrix}
\cM \\
\overline{\cM}
\end{bmatrix}
\end{equation}
where $\overline{\cM} = (\overline{m}_{ij})$ is the complement matrix of $\cM$, and $\overline{m}_{ij} = 1 - m_{ij}$ for $i = 1, \ldots, k$ and $j = 1, \ldots, n$. Given measurement matrix $\cA$ and a binary representation vector of $u$ defective items $\bX$ ($|\supp(\bX)| = u$), what we observe is $\bZ = \cA \otimes_u \bX$. The objective is to recover $\bY^\prime = \cM \odot \bX = (y^\prime_1, \ldots, y^\prime_k)^T$ from $\bZ$. Then $\bX$ can be recovered by using Theorem~\ref{thr:mainNonrandom} or~\ref{thr:mainNonrandom2}.

Assume that the outcome vector is $\bZ = \cA \otimes_u \bX = \begin{bmatrix}
\cM \otimes_u \bX \\
\overline{\cM} \otimes_u \bX
\end{bmatrix} = \begin{bmatrix}
\bY \\
\overline{\bY}
\end{bmatrix}$, where $\bY = \cM \otimes_u \bX = (y_1, \ldots, y_k)^T$ and $\overline{\bY} = \overline{\cM} \otimes_u \bX = (\overline{y}_1, \ldots, \overline{y}_k)^T$. Then vector $\bY^\prime = \cM \odot \bX$ is always obtained from $\bZ$ by using the following rules: i) If $y_l = 1$, then $y^\prime_l = 1$; ii) If $y_l = 0$ and $\overline{y}_l = 1$, then $y^\prime_l = 0$; and iii) If $y_l = 0$ and $\overline{y}_l = 0$, then $y^\prime_l = 1$. Therefore, vector $\bX$ can always be recovered. We denote the procedure to get $\bY^\prime$ by using these three rules as $\convert(\bY)$.

\section{Proposed scheme for non-adaptive classical complex group testing}
\label{sec:NACCmplxGT}

%The detailed proofs for correctness, complexity, and instantiations are available in Appendix~\ref{sec:appendix:NACCmplxGT}.

\subsection{Encoding procedure}
\label{sub:NACCmplxGT:enc}

Let $\cG$ and $\cA$ be an $h \times n$ $(d - r, r; z]$-disjunct matrix and a $2k \times n$ matrix as defined in~\eqref{eqn:elementaryMatrix}, respectively. On the basis of the final measurement matrix described in~\cite{bui2018efficiently} and~\cite{bui2018framework}, $\cT$ is created as follows:

\begin{equation}
\label{eqn:meausrementMatrix}
\cT = \begin{bmatrix}
\cG_{1, *} \\
\cA \times \diag(\cG_{1, *}) \\
\vdots \\
\cG_{h, *} \\
\cA \times \diag(\cG_{h, *})
\end{bmatrix}
= \begin{bmatrix}
\cG_{1, *} \\
\cM \times \diag(\cG_{1, *}) \\
\overline{\cM} \times \diag(\cG_{1, *}) \\
\vdots \\
\cG_{h, *} \\
\cM \times \diag(\cG_{h, *}) \\
\overline{\cM} \times \diag(\cG_{h, *})
\end{bmatrix}
\end{equation}

The vector observed after performing the tests given by $\cT$ is
\begin{align}
\bY = \cT \bullet \bX = \begin{bmatrix}
\test(\supp(\cG_{1, *}) \cap \supp(\bX)) \\
\cA \bullet \bX_1 \\
\vdots \\
\test(\supp(\cG_{h, *}) \cap \supp(\bX)) \\
\cA \bullet \bX_h
\end{bmatrix}
= \begin{bmatrix}
\test(\supp(\bX_1)) \\
\cM \bullet \bX_1  \\
\overline{\cM} \bullet \bX_1 \\
\vdots \\
\test(\supp(\bX_h)) \\
\cM \bullet \bX_h \\
\overline{\cM} \bullet \bX_h \\
\end{bmatrix}
= \begin{bmatrix}
y_1 \\
\bY_1 \\
\overline{\bY}_1 \\
\vdots \\
y_h \\
\bY_h \\
\overline{\bY}_h
\end{bmatrix}
= \begin{bmatrix}
y_1 \\
\bZ_1 \\
\vdots \\
y_h \\
\bZ_h
\end{bmatrix} \nonumber% \label{eqn:encoding}
\end{align}
where $\bX_i = \diag(\cG_{i, *}) \times \bX$, $y_i = \test(\supp(\bX_i))$, $\bY_i = \cM \bullet \bX_i = (y_{i1}, \ldots, y_{ik})^T$, $\overline{\bY}_i = \overline{\cM} \bullet \bX_i = (\overline{y}_{i1}, \ldots, \overline{y}_{ik})^T$, and $\bZ_i = [\bY_i^T \ \overline{\bY}_i^T]^T$ for $i = 1, 2, \ldots, h$.

Vector $\bX_i$ is the vector representing the defective items in row $\cG_{i, *}$. Therefore, we have $|\supp(\bX_i)| \leq d$ and $y_i = 1$ if and only if $D_a \subseteq \supp(\bX_i) = \supp(\cG_{i, *}) \cap \supp(\bX)$ for some $a \in [s]$. Once $D_a \equiv \supp(\bX_i)$, we have $y_i = \cG_{i, *} \otimes_{|D_a|} \bX$, $\bY_i = \cM \otimes_{|D_a|} \bX_i$ and $\overline{\bY}_i = \overline{\cM} \otimes_{|D_a|} \bX_i$.

\subsection{Decoding procedure}
\label{sub:NACCmplxGT:decoding}

The decoding procedure is summarized as Algorithm~\ref{alg:decodingCCmplxGT}, where $\bY^\prime_i$ is presumed to be $\cM \odot \bX_i$. The procedure is explained as follows: Step~\ref{alg:decodingCCmplxGT:scan} enumerates the $h$ rows of $\cG$. Step~\ref{alg:decodingCCmplxGT:checkPositive} checks if there is at least one positive complex in row $G_{i, *}$. Steps~\ref{alg:decodingCCmplxGT:convert} and~\ref{alg:decodingCCmplxGT:decodeAll} calculate $\bY^\prime_i$ and try to recover $\bX_i$. Step~\ref{alg:decodingCCmplxGT:sanitization} checks if all items from Step~\ref{alg:decodingCCmplxGT:decodeAll} form a positive complex. Finally, Step~\ref{alg:decodingCCmplxGT:defectiveSet} returns all positive complexes.

\begin{algorithm}
\caption{Decoding procedure for classical complex group testing}
\label{alg:decodingCCmplxGT}
\textbf{Input:} Outcome vector $\bY$, $\cM$.\\
\textbf{Output:} Positive complexes.

\begin{algorithmic}[1]
\For {$i=1$ to $h$} \label{alg:decodingCCmplxGT:scan}
	\If {$y_i = 1$} \label{alg:decodingCCmplxGT:checkPositive}
		\State $\bY^\prime_i = \convert(\bZ_i)$. \label{alg:decodingCCmplxGT:convert}
		\State $G_i = \decode(\cM, \bY^\prime_i)$. \label{alg:decodingCCmplxGT:decodeAll}
		\State \textbf{If} {$\bigwedge_{j \in G_i} \cA_j \not\equiv \bZ_i$} \textbf{then} $G_i = \emptyset$. \textbf{end if} \label{alg:decodingCCmplxGT:sanitization}
	\EndIf	
\EndFor \label{alg:decodingCCmplxGT:scan:end}
\State Return all non-emptysets $G_i$ in which its total frequency is more than $\lfloor (z-1)/2 \rfloor$. \label{alg:decodingCCmplxGT:defectiveSet}
\end{algorithmic}
\end{algorithm}

\subsection{Decoding complexity}
\label{sub:NACCmplxGT:complexity}

We summarize Algorithm~\ref{alg:decodingCCmplxGT} in the following theorem:
\begin{theorem}
\label{thr:CCmplxGT}
Let $1 \leq r \leq d < n$ and $1 \leq z, s$ be integers. Suppose that $D = \{ D_1, \ldots, D_s \}$ is an unknown set of positive complexes of $n$ items in CCmplxGT as in Definition~\ref{def:CCmplxGT}. Let $\cG$ be an $h \times n$ $(d - r, r; z]$-disjunct matrix. Suppose that a $k \times n$ $d$-disjunct matrix $\cM$ can be decoded in time $O(\sfA)$ and that each column of $\cM$ can be generated in time $O(\sfB)$. A $(2k + 1)h \times n$ measurement matrix $\cT$, as defined in~\eqref{eqn:meausrementMatrix}, can thus be used to identify all positive complexes in time $O(h(\sfA + d\sfB))$ in the presence of up to $\lfloor (z - 1)/2 \rfloor$ erroneous outcomes.
\end{theorem}

\subsection{Instantiations of decoding complexity}
\label{sub:NACCmplxGT:instan}

We instantiate Theorem~\ref{thr:CCmplxGT} by choosing $\cG$ as a $(d - r, r; z]$-disjunct matrix in Theorem~\ref{thr:ChenUpper} and $\cM$ as a $d$-disjunct matrix in Theorem~\ref{thr:mainNonrandom}.

\begin{corollary}
\label{cor:NACCmplxGT:1}
Let $1 \leq r \leq d < n$ and $1 \leq z, s$ be integers. Suppose that $D = \{ D_1, \ldots, D_s \}$ is an unknown set of positive complexes of $n$ items in CCmplxGT as in Definition~\ref{def:CCmplxGT}. There exists a $t \times n$ measurement matrix such that $D$ can be identified with $t = O \left( z \left( \frac{d}{r} \right)^r \left( \frac{d}{d - r} \right)^{d - r} d \ln{\frac{n}{d}} \cdot \frac{d^2  \ln^2{n}}{\sfW^2(d\ln{n})} \right)$ tests in time $O \left( z \left( \frac{d}{r} \right)^r \left( \frac{d}{d - r} \right)^{d - r} d \ln{\frac{n}{d}} \right) \times \sfA_0$ in the presence of up to $\lfloor (z - 1)/2 \rfloor$ erroneous outcomes, where $\sfA_0$ is defined in Table~\ref{tbl:comparison}.
\end{corollary}

To reduce the decoding complexity, matrix $\cM$ is chosen as a $d$-disjunct matrix in Theorem~\ref{thr:mainNonrandom2}. We then obtain the following corollary, in which the number of tests is larger than that in Corollary~\ref{cor:NACCmplxGT:1}.

\begin{corollary}
\label{cor:NACCmplxGT:2}
Let $1 \leq r \leq d < n$ and $1 \leq z, s$ be integers. Suppose that $D = \{ D_1, \ldots, D_s \}$ is an unknown set of positive complexes of $n$ items in CCmplxGT as in Definition~\ref{def:CCmplxGT}. There exists a $t \times n$ measurement matrix such that $D$ can be identified with $t = O \left( z \left( \frac{d}{r} \right)^r \left( \frac{d}{d - r} \right)^{d - r} d \ln{\frac{n}{d}} \cdot \frac{d^2 \ln^3{n}}{\sfW^2(d\ln{n})} \right)$ tests in time $O \left( t \times \frac{\ln{n}}{\sfW(d\ln{n})} \right)$ in the presence of up to $\lfloor (z - 1)/2 \rfloor$ erroneous outcomes.
\end{corollary}

\section{Proposed scheme for non-adaptive generalized complex group testing}
\label{sec:NAGCmplxGT}

\subsection{Encoding procedure}
\label{sub:NAGCmplxGT:enc}

Let $\cG$ and $\cA$ be an $h \times n$ $(d - u, u; z]$-disjunct matrix and a $2k \times n$ matrix as defined in~\eqref{eqn:elementaryMatrix}, respectively. Measurement matrix $\cT$ and outcome vector $\bY$ are created as in Section~\ref{sub:NACCmplxGT:enc}.

\subsection{Decoding procedure}
\label{sub:NAGCmplxGT:decoding}

The decoding procedures is summarized as Algorithm~\ref{alg:decodingGCmplxGT}. There are two phases in general: one is to identify the defective set $D^\star$ (though positive complexes are not identified), and the other one is to identify positive complexes, i.e., identify $D_a$ for all $a \in [s]$. The first phase comprises Steps~\ref{alg:decodingGCmplxGT:init} to~\ref{alg:decodingGCmplxGT:Sanitization2}. The second phase comprises Steps~\ref{alg:decodingGCmplxGT:identifyComplex_Start} to~\ref{alg:decodingGCmplxGT:defectiveSet}. Due to the length limitation, the explanation if the decoding procedure is given in Appendix~\ref{sub:appendix:NAGCmplxGT:decoding}.

\begin{algorithm}
\caption{Decoding procedure for generalized complex group testing}
\label{alg:decodingGCmplxGT}
\textbf{Input:} Outcome vector $\bY$, $\cM$.\\
\textbf{Output:} Set of defective items $S$.

\begin{algorithmic}[1]
\State $D^\star = \emptyset$. \label{alg:decodingGCmplxGT:init}
\For {$i=1$ to $h$} \label{alg:decodingGCmplxGT:scan}
	\If {$y_i = 1$} \label{alg:decodingGCmplxGT:checkPositive}
		\State $\bY^\prime_i = \convert(\bZ_i)$. \label{alg:decodingGCmplxGT:convert}
		\State $G_i = \decode(\cM, \bY^\prime_i)$. \label{alg:decodingGCmplxGT:decodeAll}
		\If {$\bigwedge_{j \in G_i} \cA_j \equiv \bZ_i$} \label{alg:decodingGCmplxGT:sanitization1}
			\State $D^\star = D^\star \cup \{G_i\}$. \label{alg:decodingGCmplxGT:add}
		\EndIf \label{alg:decodingGCmplxGT:sanitization2}
%		\State \textbf{If} \label{alg:decodingGCmplxGT:sanitization1}
	\EndIf	
\EndFor \label{alg:decodingGCmplxGT:scan:end}
\State Remove any subset in $D^\star$ appearing up to $\lfloor (z-1)/2 \rfloor$ times. \label{alg:decodingGCmplxGT:Sanitization1}
\State For a subset in $D^\star$, remove the duplicate subsets and retain the original one. \label{alg:decodingGCmplxGT:Sanitization2}
\State Distribute all subsets of $D^\star$ which have the same cadinality into a set. \label{alg:decodingGCmplxGT:identifyComplex_Start}
\State Denote these sets as $C_1, \ldots, C_v$, where $C_i = \{G_{i1}, \ldots, G_{i c_i} \}$ and $c_i = |C_i|$ for $i = 1, \ldots, v$. \label{alg:decodingGCmplxGT:subsets}
\For {$i = 1$ to $v$} \label{alg:decodingGCmplxGT:identifyComplex_startScan}
	\State $\mathrm{flag} = 0$; \label{alg:decodingGCmplxGT:identifyComplex:flag}
	\State $C_{\mathrm{new}} = \emptyset$. \label{alg:decodingGCmplxGT:identifyComplex:newSet}
	\For {$j_1 = 1$ to $|C_i| - 1$} \label{alg:decodingGCmplxGT:identifyComplex:scan1}
		\For {$j_2 = j_1 + 1$ to $|C_i|$} \label{alg:decodingGCmplxGT:identifyComplex:scan2}
			\If {$|G_{i j_1} \cap G_{i j_2}| > 0$ and $(G_{i j_1} \cup G_{i j_2}) \not\subseteq C_{i^\prime}, \forall i^\prime < i$} \label{alg:decodingGCmplxGT:identifyComplex:condition}
				\State Let $f \in G_{i j_2} \cap G_{i j_1}$. \label{alg:decodingGCmplxGT:identifyComplex:testSet}
				\While {$w \in G_{i j_1} \setminus G_{i j_2}$} \label{alg:decodingGCmplxGT:identifyComplex:While}
					\If {$((G_{i j_2} \setminus \{ f \} ) \cup w ) \not\in C_i$} \label{alg:decodingGCmplxGT:identifyComplex:checkBelonging}
						\State $\mathrm{flag} = 1$. \label{alg:decodingGCmplxGT:identifyComplex:TurnOnFlag}
						\State $C_{\mathrm{new}} = C_{\mathrm{new}} \cup \{ G_{i j_2} \}$. \label{alg:decodingGCmplxGT:identifyComplex:add}
						\State $C_i = C_i \setminus G_{i j_2}$. \label{alg:decodingGCmplxGT:identifyComplex:remove}
						\State $j_2 = j_2 - 1$; \label{alg:decodingGCmplxGT:identifyComplex:correct}
					\EndIf
				\EndWhile \label{alg:decodingGCmplxGT:identifyComplex:testSet_End}
			\EndIf
		\EndFor
	\EndFor
	\If {$\mathrm{flag} = 1$} \label{alg:decodingGCmplxGT:identifyComplex:newCluster_Start}
		\State $v = v + 1$;
		\State $C_v = C_{\mathrm{new}}$.
	\EndIf \label{alg:decodingGCmplxGT:identifyComplex:newCluster_End}
%	\State \textbf{If} \label{alg:decodingGCmplxGT:identifyComplex:newCluster_Start}
	\State $C_i = \plain(C_i)$. \label{alg:decodingGCmplxGT:identifyComplex}
\EndFor \label{alg:decodingGCmplxGT:identifyComplex_endScan}
\State Return all $C_i$s. \label{alg:decodingGCmplxGT:defectiveSet}
\end{algorithmic}
\end{algorithm}

\subsection{Decoding complexity}
\label{sub:NAGCmplxGT:complexity}

We summarize Algorithm~\ref{alg:decodingGCmplxGT} in the following theorem:
\begin{theorem}
\label{thr:GCmplxGT}
Let $1 \leq r \leq d < n$ and $1 \leq z, s$ be integers. Suppose that $D = \{ D_1, \ldots, D_s \}$ is an unknown set of positive complexes of $n$ items in GCmplxGT as in Definition~\ref{def:GCmplxGT}. Let $\cG$ be an $h \times n$ $(d - u, u; z]$-disjunct matrix. Suppose that a $k \times n$ $d$-disjunct matrix $\cM$ can be decoded in time $O(\sfA)$ and that each column of $\cM$ can be generated in time $O(\sfB)$. A $(2k + 1)h \times n$ measurement matrix $\cT$, as defined in~\eqref{eqn:meausrementMatrix}, can thus be used to identify all positive complexes in time $O(h(\sfA + d\sfB) + su^3dq^3)$ in the presence of up to $\lfloor (z - 1)/2 \rfloor$ erroneous outcomes, where $q = \sum_{a = 1}^s \binom{|D_a|}{u_a}$. When $u_x \neq u_y$ for any $x \neq y \in [s]$, the term $su^3dq^3$ can be removed.
\end{theorem}

\subsection{Instantiations of decoding complexity}
\label{sub:NAGCmplxGT:instan}

We instantiate Theorem~\ref{thr:GCmplxGT} by choosing $\cG$ as a $(d - u, u; z]$-disjunct matrix in Theorem~\ref{thr:ChenUpper} and $\cM$ as a $d$-disjunct matrix in Theorem~\ref{thr:mainNonrandom}.

\begin{corollary}
\label{cor:NAGCmplxGT:1}
Let $1 \leq r \leq d < n$ and $1 \leq z, s$ be integers. Suppose that $D = \{ D_1, \ldots, D_s \}$ is an unknown set of positive complexes of $n$ items in GCmplxGT as in Definition~\ref{def:GCmplxGT}. There exists a $t \times n$ measurement matrix such that $D$ can be identified with $t = O \left( z \left( \frac{d}{u} \right)^u \left( \frac{d}{d - u} \right)^{d - u} d \ln{\frac{n}{d}} \cdot \frac{d^2  \ln^2{n}}{\sfW^2(d\ln{n})} \right)$ tests in time $O \left( z \left( \frac{d}{u} \right)^u \left( \frac{d}{d - u} \right)^{d - u} d \ln{\frac{n}{d}} \right) \times \sfA_0 + O(su^3dq^3)$ in the presence of up to $\lfloor (z - 1)/2 \rfloor$ erroneous outcomes, where $q = \sum_{a = 1}^s \binom{|D_a|}{u_a}$ ($\sfA_0$ is defined in Table~\ref{tbl:comparison}). When $u_x \neq u_y$ for any $x \neq y \in [s]$, the term $su^3dq^3$ can be removed.
\end{corollary}

To reduce decoding complexity, matrix $\cM$ is chosen as a $d$-disjunct matrix in Theorem~\ref{thr:mainNonrandom2}. We then obtain the following corollary, in which the number of tests is larger than in Corollary~\ref{cor:NAGCmplxGT:1}.

\begin{corollary}
\label{cor:NAGCmplxGT:2}
Let $1 \leq r \leq d < n$ and $1 \leq z, s$ be integers. Suppose that $D = \{ D_1, \ldots, D_s \}$ is an unknown set of positive complexes of $n$ items in GCmplxGT as in Definition~\ref{def:GCmplxGT}. There exists a $t \times n$ measurement matrix such that $D$ can be identified with $t = O \left( z \left( \frac{d}{u} \right)^u \left( \frac{d}{d - u} \right)^{d - u} d \ln{\frac{n}{d}} \cdot \frac{d^2 \ln^3{n}}{\sfW^2(d\ln{n})} \right)$ tests in time $O \left( t \times \frac{\ln{n}}{\sfW(d\ln{n})} \right) + O(su^3dq^3)$ in the presence of up to $\lfloor (z - 1)/2 \rfloor$ erroneous outcomes, where $q = \sum_{a = 1}^s \binom{|D_a|}{u_a}$. When $u_x \neq u_y$ for any $x \neq y \in [s]$, the term $su^3dq^3$ can be removed.
\end{corollary}

%\section{Conclusion}
%\label{sec:cls}
%
%We have proposed efficient and error-tolerant schemes for both CCmplxGT and GCmplxGT with non-adaptive design, and showed the connection between CmplxGT and complex disease genetics. The proposed scheme for CCmplxGT is much better than previous schemes in terms of decoding time. Moreover, when $s > d$, the number of tests in our proposed scheme is also smaller than the one in previous work. The proposed scheme for GCmplxGT is the first one proposed. Both schemes mainly rely on the construction of disjunct matrices, which increase the number of tests and decoding time. Since we only need certain rows in the disjunct matrices that have specific properties, we should be able to dramatically reduce the number of tests and decoding time by generating matrices with rows having those properties. Doing this remains for future work.
%
%
\section{Acknowledgments}

The first author thanks Dr. Nikita Polyanskii, Skolkovo Institute of Science and Technology, Russia, for his discussion on hidden hypergraphs.

%\vspace{-1mm}
\bibliographystyle{ieeetr}
\bibliography{bibli}
%\vspace{-2mm}

\appendix

\section{Omitted proofs in Section~\ref{sec:NACCmplxGT}}
\label{sec:appendix:NACCmplxGT}

\subsection{Correctness of the decoding procedure}
\label{sub:NACCmplxGT:correctness}

Our objective is to recover $\bX_i$ from $y_i$ and $\bZ_i = \begin{bmatrix} \bY_i \\ \overline{\bY}_i \end{bmatrix}$ for $i = 1, \ldots, h$. We pay attention only to the $\bX_i$ in which $\supp(\bX_i) \equiv D_a$ for some $a \in [s]$ because otherwise $\supp(\bX_i)$ should not be counted as a positive complex.

Step~\ref{alg:decodingCCmplxGT:scan} enumerates the $h$ rows of $\cG$. We have that $y_i$ is the indicator for whether there is at least one positive complex in row $\cG_{i, *}$. Because we focus only on rows $\cG_{i, *}$ that have exactly one positive complex, vector $\bZ_i$ is not considered if $y_i = 0$. This is achieved by Step~\ref{alg:decodingCCmplxGT:checkPositive}.

When $y_i = 1$, there is at least one positive complex in row $\cG_{i, *}$. If there is only one positive complex in row $\cG_{i, *}$, say $D_a$, then $G_i \equiv D_a$ for the reason described in the last paragraph of Section~\ref{sub:NACCmplxGT:enc} and the scheme in Section~\ref{sec:review}. Therefore, the condition in Step~\ref{alg:decodingCCmplxGT:sanitization} does not hold. As a result, the positive complex $G_i$ will not be empty after Step~\ref{alg:decodingCCmplxGT:sanitization}. Because $\cG$ is a $(d -r, r; z]$, each $D_a$ is identical to the support set of at least $z$ rows of $\cG$ for $a \in [s]$. In other words, there exists at least $z$ rows $\cG_{i, *}$s such that $\supp(\cG_{i, *}) \cap D^\prime = D_a$ for every $a \in [s]$. Therefore, each $D_a$ will be returned more than $\lfloor (z-1)/2 \rfloor$ times if there are up to $\lfloor (z - 1)/2 \rfloor$ erroneous outcomes.

The following argument considers the case in which there are at least two positive complexes present in row $\cG_{i, *}$. Our task is now to prevent false defectives, i.e., to prove that Step~\ref{alg:decodingCCmplxGT:sanitization} removes all false defective items. There are two sets of defectives items corresponding to $\bZ_i$: the first one is the true set, which is $S_i = \supp(\cG_{i, *}) \cap D^\prime$ and is unknown; the second one is the recovered set $G_i$, which is expected to be $S_i$ (although not surely). We consider two cases: $|G_i \setminus D^\prime| = 0$ and $|G_i \setminus D^\prime| > 0$, where $D^\prime = \plain(D)$. For the latter case, we prove that the condition $\bigwedge_{j \in G_i} \cA_j \not\equiv \bZ_i$ holds. Consider item $j_0 \in G_i \setminus D^\prime$. Because $\cM$ is a $d$-disjunct matrix, there exists row $\tau$ such that $m_{\tau j_0} = 1$ and $m_{\tau j} = 0$ for all $j \in D^\prime$. Hence, $\overline{m}_{\tau j_0} = 0$ and $\overline{m}_{\tau j} = 1$ for all $j \in D^\prime$. It follows that $\overline{y}_{i \tau} = 1$ because $\overline{m}_{\tau j} = 1$ for all $j \in D^\prime$ and $D_a \subseteq \supp(\overline{\cM}_{\tau, *}) \cap D^\prime$ for some $a \in [s]$. However, $\bigwedge_{j \in G_i} \overline{m}_{\tau j} = \left( \bigwedge_{j \in G_i \setminus \{j_0 \} } \overline{m}_{\tau j} \right) \wedge \overline{m}_{\tau j_0} = \left( \bigwedge_{j \in G_i \setminus \{j_0 \} } \overline{m}_{\tau j} \right) \wedge 0 = 0 \neq \overline{y}_{i \tau} = 1$. Therefore, the condition at Step~\ref{alg:decodingCCmplxGT:sanitization} holds, i.e., the set $G_i$ is set to be an empty set and not be counted as a positive complex.

We consider the remaining case when $|G_i \setminus D^\prime| = 0$. In this case, we also prove that the condition at Step~\ref{alg:decodingCCmplxGT:sanitization} holds. Indeed, let $D_a$ and $D_b$ be two positive complexes in row $\cG_{i, *}$. Although there may be more than two positive complexes in row $\cG_{i, *}$, we consider only two of them. Let $j_1$ be an item in $D_a$. Because $\cM$ is a $d$-disjunct matrix, there exists row $\chi$ such that $m_{\chi j_1} = 1$ and $m_{\chi j} = 0$ for all $j \in D_b$. Hence, $\overline{m}_{\chi j_1} = 0$ and $\overline{m}_{\chi j} = 1$ for all $j \in D_b$. Then $\overline{y}_{i \chi} = 1$ because $\overline{m}_{\chi j} = 1$ for all $j \in D_b$ and $D_b \subseteq \supp(\overline{\cM}_{\chi, *}) \cap D^\prime$. However, $\bigwedge_{j \in G_i} \overline{m}_{\chi j} = \left( \bigwedge_{j \in G_i \setminus \{j_1 \} } \overline{m}_{\chi j} \right) \wedge \overline{m}_{\chi j_1} = \left( \bigwedge_{j \in G_i \setminus \{j_1 \} } \overline{m}_{\chi j} \right) \wedge 0 = 0 \neq \overline{y}_{i \chi} = 1$. Therefore, if there are at least two positive complexes in row $\cG_{i, *}$, the condition at Step~\ref{alg:decodingCCmplxGT:sanitization} holds; i.e., set $G_i$ will be set to be an empty set and not be counted as a positive complex.

In conclusion, Algorithm~\ref{alg:decodingCCmplxGT} returns all positive complexes with up to $(z - 1)/2$ erroneous outcomes.

\subsection{Decoding complexity}
\label{sub:appendix:NACCmplxGT:complexity}

The time to run Step~\ref{alg:decodingCCmplxGT:convert} is $O(k)$. Suppose that matrix $\cM$ can be decoded in time $O(\sfA)$ and that each column in $\cM$ can be generated in time $O(\sfB)$. It is natural that $k \leq O(A), O(B)$ because each column in $\cM$ has $k$ entries. It thus takes $O(\sfA)$ time to rune Step~\ref{alg:decodingCCmplxGT:decodeAll}. Since $\cM$ is $d$-disjunct, the cardinality of any $G_i$ obtained in Step~\ref{alg:decodingCCmplxGT:sanitization} is not exceeded $d$. Therefore, it takes $d \times O(\sfB)$ time to run Step~\ref{alg:decodingCCmplxGT:sanitization}. Step~\ref{alg:decodingCCmplxGT:defectiveSet} takes $O(d \times h)$ time to run. Because the loop in Step~\ref{alg:decodingCCmplxGT:scan} runs $h$ times, the total decoding time is:
\begin{equation}
h \times (O(k) + \sfA + d \times O(\sfB)) + O(dh) = O(h(\sfA + d\sfB)).
\end{equation}

\subsection{Instantiations of decoding complexity}
\label{sub:appendix:NACCmplxGT:instan}

With the setting in Corollary~\ref{cor:NACCmplxGT:1}, we have:
\begin{align}
h &= O \left( z \left( \frac{d}{r} \right)^r \left( \frac{d}{d - r} \right)^{d - r} d \ln{\frac{n}{d}} \right), &k &= O \left(\frac{d^2 \ln^2{n}}{\sfW^2(d\ln{n})} \right), \nonumber \\
\sfA &= O \left( \frac{d^{3.57} \ln^{6.26}{n}}{\sfW^{6.26}(d \ln{n})} \right) + O \left( \frac{d^6 \ln^4{n}}{\sfW^4(d \ln{n})} \right), &\sfB &= O \left( \frac{k^{1.5}}{d^2} \right) = O \left( \frac{d \ln^3{n}}{\sfW^3(d\ln{n})} \right). \nonumber
\end{align}

\noindent
For Corollary~\ref{cor:NACCmplxGT:2}, we have:

\begin{align}
h &= O \left( z \left( \frac{d}{r} \right)^r \left( \frac{d}{d - r} \right)^{d - r} d \ln{\frac{n}{d}} \right), &k &= O \left( \frac{d^2 \ln^3{n}}{\sfW^2(d \ln{n})} \right), \nonumber \\
\sfA &= O \left( \frac{d^2 \ln^3{n}}{\sfW^2(d \ln{n})} \right), &\sfB &= O \left(\frac{d \ln^4{n}}{\sfW^3(d\ln{n})} \right). \nonumber
\end{align}

\section{Omitted proofs in Section~\ref{sec:NAGCmplxGT}}
\label{sec:appendix:NAGCmplxGT}

\subsection{Decoding procedure}
\label{sub:appendix:NAGCmplxGT:decoding}

As in Algorithm~\ref{alg:decodingCCmplxGT}, Step~\ref{alg:decodingGCmplxGT:scan} in the first phase enumerates the $h$ rows of $\cG$. Step~\ref{alg:decodingGCmplxGT:checkPositive} checks if there is at least one positive complex in row $G_{i, *}$. Step~\ref{alg:decodingGCmplxGT:convert} calculates $\bY^\prime_i$, which is presumed to be $\cM \odot \bX_i$. Step~\ref{alg:decodingGCmplxGT:decodeAll} then tries to recover $\bX_i$. Steps~\ref{alg:decodingGCmplxGT:sanitization1} to~\ref{alg:decodingGCmplxGT:sanitization2} accept only a positive sub-complex in which all the elements belong to some positive complex and the cardinality of the sub-complex equals the threshold of the positive complex. Step~\ref{alg:decodingGCmplxGT:Sanitization1} removes the false positive complexes that may appear after running the previous steps. The final step in the first phase lists all positive sub-complexes without duplicates in Step~\ref{alg:decodingGCmplxGT:Sanitization2}. It follows that $D^\star$ becomes a set after this step.

The second phase is to identify positive complexes. As a result of the steps in the first phase, all elements in each subset in $D^\star$ belong to some positive complex and the cardinality of the subset equals the threshold of the corresponding positive complex. In the second phase, $D^\star$ is first partitioned on the basis of the cardinalities of its subsets, as described in Step~\ref{alg:decodingGCmplxGT:identifyComplex_Start}. Each partition is called \textit{a set}. Assume there are $v$ sets as in Step~\ref{alg:decodingGCmplxGT:subsets}. When $u_x \neq u_y$ for any $x \neq y \in [s]$, Steps~\ref{alg:decodingGCmplxGT:identifyComplex_startScan} to~\ref{alg:decodingGCmplxGT:identifyComplex_endScan} can be replaced by Step~\ref{alg:decodingGCmplxGT:identifyComplex}.

Step~\ref{alg:decodingGCmplxGT:identifyComplex_startScan} scans $v$ sets to identify positive complexes. Since two positive complexes could share the same threshold, a set may contain at least two positive complexes. We thus create a flag in Step~\ref{alg:decodingGCmplxGT:identifyComplex:flag} to indicate whether this happens. If there are at least two positive complexes in a set, our objective is to keep all positive sub-complexes in which the elements belong to a positive complex in the original set and to move the remaining positive sub-complexes into a new set $C_{\mathrm{new}}$. Step~\ref{alg:decodingGCmplxGT:identifyComplex:newSet} declares variable $C_{\mathrm{new}}$ for this case case.

For each set, two subsets, say $A$ and $B$, having a common element fall into two categories: they are two positive sub-complexes of a positive complex or they are two positive sub-complexes of two positive complexes. Steps~\ref{alg:decodingGCmplxGT:identifyComplex:scan1} to~\ref{alg:decodingGCmplxGT:identifyComplex:scan2} scan every pair of subsets in the set. Step~\ref{alg:decodingGCmplxGT:identifyComplex:condition} checks whether the two subsets intersects and whether they are subsets of any set formed in the previous loops, i.e., $C_1, \ldots, C_{i-1}$ if we are considering $C_i$. A new subset is created by removing an element in $A \cap B$ and adding an element in $A \setminus B$, as described in Steps~\ref{alg:decodingGCmplxGT:identifyComplex:testSet} to~\ref{alg:decodingGCmplxGT:identifyComplex:checkBelonging}. Step~\ref{alg:decodingGCmplxGT:identifyComplex:checkBelonging} validates the second category by checking whether the new subset belongs to the set. The indicator for having at least two positive complexes in the set is thus turned on at Step~\ref{alg:decodingGCmplxGT:identifyComplex:TurnOnFlag}. Subset $B$ is then added to the new set $C_{\mathrm{new}}$ in Step~\ref{alg:decodingGCmplxGT:identifyComplex:add} and removed from the current set in Step~\ref{alg:decodingGCmplxGT:identifyComplex:remove}. Step~\ref{alg:decodingGCmplxGT:identifyComplex:correct} adjusts the loop at Step~\ref{alg:decodingGCmplxGT:identifyComplex:scan2} due tos the change in sets in Step~\ref{alg:decodingGCmplxGT:identifyComplex:remove}. If the flag is turned on, Steps~\ref{alg:decodingGCmplxGT:identifyComplex:newCluster_Start} to~\ref{alg:decodingGCmplxGT:identifyComplex:newCluster_End} adjust the sets accordingly. Step~\ref{alg:decodingGCmplxGT:identifyComplex} merges all positive sub-complexes to form a positive complex. Finally, Step~\ref{alg:decodingGCmplxGT:defectiveSet} returns all positive complexes.

\subsection{Correctness of decoding procedure}
\label{sub:appendix:NAGCmplxGT:correctness}

Recall that there are two phases in general: one to identify the defective set $D^\star$ (though positive complexes are not identified), and the other to identify positive complexes, i.e., identify $D_a$ for $a \in [s]$. The first and second phases consist of Steps~\ref{alg:decodingGCmplxGT:init} to~\ref{alg:decodingGCmplxGT:Sanitization2} and Steps~\ref{alg:decodingGCmplxGT:identifyComplex_Start} to~\ref{alg:decodingGCmplxGT:defectiveSet}, respectively. We move to the first phase in details now.

\subsubsection{Locating defective items}
\label{subsub:NAGCmplxGT:locate}

We first assume that there are no erroneous outcomes. Our objective is to recover $\bX_i$ from $y_i$ and $\bZ_i = \begin{bmatrix} \bY_i \\ \overline{\bY}_i \end{bmatrix}$ for $i = 1, \ldots, h$. We recover only $\bX_i$ if $|\supp(\bX_i) \setminus D_a| = 0$ and $|\supp(\bX_i)| = u_a$ for some $a \in [s]$. For this condition, we make \textit{an ideal assumption} that, for every $\bX_i$, there exists only $D_a$ such that $|\supp(\bX_i) \setminus D_a| = 0$ and $|\supp(\bX_i)| = u_a$ for some $a \in [s]$. If this assumption does not hold, $\supp(\bX_i)$ is not added to $D^\star$.

Step~\ref{alg:decodingCCmplxGT:scan} enumerates the $h$ rows of $\cG$. We have that $y_i$ is the indicator for whether there is at least one positive sub-complex in row $\cG_{i, *}$. If $y_i = 0$, $\supp(\bX_i)$ is an incomplete positive sub-complex, so $\bZ_i$ is not considered. This is done by running Step~\ref{alg:decodingGCmplxGT:checkPositive}.

We now consider the case $y_i = 1$; i.e., there is at least one positive sub-complex in row $\cG_{i, *}$. There are three possibilities:
\begin{itemize}
\item There is only one positive sub-complex in row $\cG_{i, *}$ such that $|\supp(\bX_i) \setminus D_a| = 0$ and $|\supp(\bX_i)| = u_a$ for some $a \in [s]$. This is the ideal assumption.
\item There is only one positive sub-complex in row $\cG_{i, *}$ such that $|\supp(\bX_i) \setminus D_a| = 0$ and $|\supp(\bX_i)| > u_a$ for some $a \in [s]$.
\item There are more than two positive sub-complexes in row $\cG_{i, *}$.
\end{itemize} 

There are two sets of defectives items corresponding to $\bZ_i$: the first one is the true set, which is $S_i = \supp(\bX_i)$ and is unknown, and the second one is the recovered set $G_i$, which is expected to be $S_i$ (although not surely). Since we made the ideal assumption, Steps~\ref{alg:decodingGCmplxGT:convert} and~\ref{alg:decodingGCmplxGT:decodeAll} are simply to implement the procedure described in Section~\ref{sec:review}. The important point is that $G_i \equiv \supp(\bX_i)$ if the ideal assumption holds without knowing the exact value of $u_a$ (as long as $u_a \leq d$). Step~\ref{alg:decodingGCmplxGT:sanitization1} checks whether the ideal assumption holds. If the first possibility occurs, it is obvious that the ideal assumption holds. Step~\ref{alg:decodingGCmplxGT:add} is thus implemented.

We now prove that when the second or third possibility occurs, i.e., the ideal assumption does not hold, the condition in Step~\ref{alg:decodingGCmplxGT:add} does not hold. Consequently, there is no positive sub-complex to be added to the defective set $D^\star$.

Consider the second possibility. We break it down into two cases: $|G_i \setminus D^\prime| > 0$ and $|G_i \setminus D^\prime| = 0$, where $D^\prime = \plain(D)$. In the first case, the argument is similar to the one in Section~\ref{sub:NACCmplxGT:correctness}. Consider item $j_0 \in G_i \setminus D^\prime$. Because $\cM$ is a $d$-disjunct matrix, there exists row $\tau$ such that $m_{\tau j_0} = 1$ and $m_{\tau j} = 0$ for all $j \in D^\prime$. Hence, $\overline{m}_{\tau j_0} = 0$ and $\overline{m}_{\tau j} = 1$ for all $j \in D^\prime$. It follows that $\overline{y}_{i \tau} = 1$ because $D_a \subseteq \supp(\bX_i)$ for some $a \in [s]$ and $\overline{m}_{\tau j} = 1$ for all $j \in D^\prime$. However, $\bigwedge_{j \in G_i} \overline{m}_{\tau j} = \left( \bigwedge_{j \in G_i \setminus \{j_0 \} } \overline{m}_{\tau j} \right) \wedge \overline{m}_{\tau j_0} = \left( \bigwedge_{j \in G_i \setminus \{j_0 \} } \overline{m}_{\tau j} \right) \wedge 0 = 0 \neq \overline{y}_{i \tau} = 1$. Therefore, the condition at Step~\ref{alg:decodingGCmplxGT:sanitization1} does not hold.

We now consider the remaining case in which $|G_i \setminus D^\prime| = 0$. Using the same argument as above, we consider item $j_0 \in G_i$. Because $\cM$ is a $d$-disjunct matrix and $|G_i| \leq d$, there exists row $\upsilon$ such that $m_{\upsilon j_0} = 1$ and $m_{\upsilon j} = 0$ for all $j \in G_i \setminus \{ j_0 \}$. Hence, $\overline{m}_{\upsilon j_0} = 0$ and $\overline{m}_{\upsilon j} = 1$ for all $j \in D^\prime$. Moreover, $D^\prime \setminus \{ j_0 \}$ must contain a positive sub-complex because of the condition in the second possibility. Therefore, $\overline{y}_{i \upsilon} = 1$. On the other hand, we have $\bigwedge_{j \in G_i} \overline{m}_{\upsilon j} = \left( \bigwedge_{j \in G_i \setminus \{j_0 \} } \overline{m}_{\upsilon j} \right) \wedge \overline{m}_{\upsilon j_0} = \left( \bigwedge_{j \in G_i \setminus \{j_0 \} } \overline{m}_{\upsilon j} \right) \wedge 0 = 0 \neq \overline{y}_{i \upsilon} = 1$. Therefore, the condition at Step~\ref{alg:decodingGCmplxGT:sanitization1} does not hold.

For the third possibility, let $D_a$ and $D_b$ be two positive sub-complexes in row $\cG_{i, *}$. Although there might be more than two positive sub-complexes in row $\cG_{i, *}$, we only consider two of them. Following the argument in section~\ref{sub:NACCmplxGT:correctness}, there always exists a row $\chi$ such that $\bigwedge_{j \in G_i} \overline{m}_{\chi j} = 0 \neq \overline{y}_{i \chi} = 1$. Therefore, the condition at Step~\ref{alg:decodingGCmplxGT:sanitization1} does not hold.

After running Steps~\ref{alg:decodingGCmplxGT:init} to~\ref{alg:decodingGCmplxGT:scan:end}, every member $M$ of $D^\star$ satisfies $|M \setminus D_a| = 0$ and $|M| = u_a$ for some $a \in [s]$. For each $D_a$, there are $\binom{|D_a|}{u_a}$ such $M$ if the frequency of $M$ in $D^\star$ is not considered. Because $\cG$ is a $(d - u, u; z]$-disjunct matrix and $u_a \leq u$, there exists at least $z$ rows $\cG_{i, *}$ such that $M = \supp(\bX_i)$.

We now consider erroneous outcomes. Since there are up to $\lfloor (z-1)/2 \rfloor$ erroneous outcomes, any false positive sub-complex in $D^\star$ cannot appear more than $\lfloor (z-1)/2 \rfloor$ times. Therefore, we can eliminate them by checking their frequencies in $D^\star$. This sanitization procedure is done by running Step~\ref{alg:decodingGCmplxGT:Sanitization1}. Finally, we keep only one copy of each positive sub-complex in $D^\star$ by running Step~\ref{alg:decodingGCmplxGT:Sanitization2}.

In summary, this phase results in a set $D^\star$ such that each member $M$ of $D^\star$ satisfies $|M \setminus D_a| = 0$ and $|M| = u_a$ for some $a \in [s]$. Moreover, for each $D_a$, the number of such $M$ is $\binom{|D_a|}{u_a}$.

\subsubsection{Identifying positive complexes}
\label{subsub:NAGCmplxGT:complex}

There are two conditions to accomplish this phase:
\begin{enumerate}
\item All elements of every subset of $D^\star$ belong to a positive complex such that the cardinality of the subset is equal to the threshold of the positive complex.
\item For any two subsets of $D^\star$, say $A$ and $B$, and $|A| = |B| = c$, if all elements of $A \cup B$ do not belong to a positive complex, there exists an element $x \in A$ and an element $y \in B$ such that $D^\star$ does not contain the set $\{ x \} \cup B \setminus \{ y \}$.
\end{enumerate}

The first condition is accomplished in the first phase. We now prove the second condition. Because of the first condition, there exists $a, b \in [s]$ such that $|A \cap D_a| = 0$, $|A| = u_a = c$, $|B \cap D_b| = 0$, and $|B| = u_b = c$. If all elements of $A \cup B$ do not belong to a positive complex, i.e., $A \cup B \not\subseteq D_a$ and $A \cup B \not\subseteq D_b$, we must have $a \neq b$ and $|A \cap B| \leq c - 1$. Therefore, there exists an element $x \in A$ and an element $y \in B$ such that the set $\{ x \} \cup (B \setminus \{ y \})$ is an incomplete positive sub-complex. Therefore, $D^\star$ does not contain the set $\{ x \} \cup (B \setminus \{ y \})$.

We now prove that the second phases consisting of Steps~\ref{alg:decodingGCmplxGT:identifyComplex_Start} to~\ref{alg:decodingGCmplxGT:defectiveSet} returns all positive complexes. Because of the first condition, all elements of a positive complex are distributed into several subsets in which the cardinality of the subset is equal to the threshold of the positive complex. Step~\ref{alg:decodingGCmplxGT:identifyComplex_Start} is thus to create sets containing subsets of positive complexes on the basis of their thresholds. After running Step~\ref{alg:decodingGCmplxGT:identifyComplex_Start}, all elements of a positive complex must belong to a set. Therefore, when $u_x \neq u_y$ for any $x \neq y \in [s]$, Steps~\ref{alg:decodingGCmplxGT:identifyComplex_startScan} to~\ref{alg:decodingGCmplxGT:identifyComplex_endScan} can be replaced by Step~\ref{alg:decodingGCmplxGT:identifyComplex}

Every set created after Step~\ref{alg:decodingGCmplxGT:identifyComplex_Start} is investigated in Step~\ref{alg:decodingGCmplxGT:identifyComplex_startScan}. For a set, says $C$, there are two possibilities: its plain set contains only a positive complex or more than two positive complexes. The second possibility can be detected by using the two conditions. The first condition provides a strategy for forming a positive complex: for a subset in $C$ there always exists another subset such that these two subsets have a common element and their union is a subset of a positive complex. With this strategy, we scan every pair of subsets in a set. This procedure is done by running Steps~\ref{alg:decodingGCmplxGT:identifyComplex:scan1} and~\ref{alg:decodingGCmplxGT:identifyComplex:scan2}. If two subsets, say $A$ and $B$, have common elements, they must be subsets of a positive complex or be two subsets of two positive complexes. It is easy to determine into which case these two subsets fall. Step~\ref{alg:decodingGCmplxGT:identifyComplex:condition} checks whether $A$ and $B$ have common elements and whether they are subsets of any set formed in previous loops, i.e., $C_1, \ldots, C_{i-1}$ if we are considering set $C_i$. Because each subset has at least one element that does not belong to the other, we create a new set $x \cup (A \setminus y)$, where $x \in A \setminus B$ and $y \in A \cap B$. Because the cardinality of the new subset is equal to $|A|$, it must belong to set $C$ if the two subsets $A$ and $B$ are subsets of a positive complex. Otherwise, the new subset does not belong to $C$ because it is an incomplete positive sub-complex. If the two subsets belong to two positive complexes, we simply place one subset into a new set and consider it later. These actions are done in Steps~\ref{alg:decodingGCmplxGT:identifyComplex:testSet} to~\ref{alg:decodingGCmplxGT:identifyComplex:testSet_End}. After these steps are run, set $C$ contains only subsets of a positive complex. Therefore, Step~\ref{alg:decodingGCmplxGT:identifyComplex} forms that positive complex.

\subsection{Decoding complexity}
\label{sub:appendix:NAGCmplxGT:complexity}

Suppose that matrix $\cM$ can be decoded in time $O(\sfA)$ and that each column in $\cM$ can be generated in time $O(\sfB)$. From the same analysis described in Section~\ref{sub:NACCmplxGT:decoding}, the complexity of Steps~\ref{alg:decodingGCmplxGT:init} to~\ref{alg:decodingGCmplxGT:Sanitization1} is $O(h(\sfA + d\sfB))$. Because $|\plain(D^\star)| \leq dh$,  Steps~\ref{alg:decodingGCmplxGT:Sanitization2} to~\ref{alg:decodingGCmplxGT:identifyComplex_Start} take $O(dh)$ time. 

From the first condition in Section~\ref{subsub:NAGCmplxGT:complex}, the cardinality of $D^\star$ is 
\begin{equation}
q = \sum_{a = 1}^s \binom{|D_a|}{u_a}.
\end{equation}

Algorithm~\ref{alg:decodingGCmplxGT} runs $v \leq s$ loops in Step~\ref{alg:decodingGCmplxGT:identifyComplex_startScan} and takes $|C_i| (|C_i| - 1)/2$ time to run the loops in Steps~\ref{alg:decodingGCmplxGT:identifyComplex:scan1} and~\ref{alg:decodingGCmplxGT:identifyComplex:scan2}. Because $u_a \leq u$ for all $a \in [s]$, Step~\ref{alg:decodingGCmplxGT:identifyComplex:condition} takes $O(ud)$ time. Steps~\ref{alg:decodingGCmplxGT:identifyComplex:While}, ~\ref{alg:decodingGCmplxGT:identifyComplex:checkBelonging}, and~\ref{alg:decodingGCmplxGT:identifyComplex} take $O(u)$ loops, $O(u|C_i|)$ time, and $O(|C_i|)$ time to run, respectively. Because $|D^\star| = q$, we have $|C_i| \leq q$. Therefore, Steps~\ref{alg:decodingGCmplxGT:identifyComplex_startScan} to~\ref{alg:decodingGCmplxGT:identifyComplex_endScan} take time
\begin{equation}
v \times \left( \frac{|C_i| (|C_i| - 1)}{2} \times ud \times u \times u|C_i| + |C_i| \right) \leq O(su^3dq^3).
\end{equation}

In summary, the total decoding time is
\begin{equation}
O(h(\sfA + d\sfB)) + O(su^3dq^3) = O(h(\sfA + d\sfB)) + O \left( su^3d \left( \sum_{a = 1}^s \binom{|D_a|}{u_a} \right)^3 \right). \nonumber
\end{equation}

\subsection{Instantiations of decoding complexity}
\label{sub:appendix:NAGCmplxGT:instan}

With the setting in Corollary~\ref{cor:NAGCmplxGT:1}, we have:
\begin{align}
h &= O \left( z \left( \frac{d}{u} \right)^u \left( \frac{d}{d - u} \right)^{d - u} d \ln{\frac{n}{d}} \right), &k &= O \left(\frac{d^2 \ln^2{n}}{\sfW^2(d\ln{n})} \right), \nonumber \\
\sfA &= O \left( \frac{d^{3.57} \ln^{6.26}{n}}{\sfW^{6.26}(d \ln{n})} \right) + O \left( \frac{d^6 \ln^4{n}}{\sfW^4(d \ln{n})} \right), &\sfB &= O \left( \frac{k^{1.5}}{d^2} \right) = O \left( \frac{d \ln^3{n}}{\sfW^3(d\ln{n})} \right). \nonumber
\end{align}

\noindent
For Corollary~\ref{cor:NAGCmplxGT:2}, we have:

\begin{align}
h &= O \left( z \left( \frac{d}{u} \right)^u \left( \frac{d}{d - u} \right)^{d - u} d \ln{\frac{n}{d}} \right), &k &= O \left( \frac{d^2 \ln^3{n}}{\sfW^2(d \ln{n})} \right), \nonumber \\
\sfA &= O \left( \frac{d^2 \ln^3{n}}{\sfW^2(d \ln{n})} \right), &\sfB &= O \left(\frac{d \ln^4{n}}{\sfW^3(d\ln{n})} \right). \nonumber
\end{align}

% that's all folks
\end{document}